\documentstyle[12pt]{article}
\let\section=\subsection     \let\subsection=\subsubsection                %%
\newcommand{\beqn}{\begin{equation}}
\newcommand{\eeqn}{\end{equation}}
\newcommand{\auau}{Au$+$Au}
\usepackage{cite}

\begin{document}
\begin{center}
   {\large \bf FLOW PHENOMENA AT AGS ENERGIES}\\[5mm]
   J.P.~WESSELS \\[5mm]
   {\small \it  Physikalisches Institut, Universit\"at Heidelberg, \\
   Philosophenweg 12, 69120 Heidelberg, Germany \\[8mm] }
\end{center}

\begin{abstract}\noindent
  In this talk some of the latest data on {\em directed sideward}, {\em
  elliptic}, {\em radial}, and {\em longitudinal} flow at AGS energies
  will be reviewed. A method to identify the reaction plane
  event by event and the measurement of its resolution will be discussed. 
  The distributions of global observables (transverse energy $E_T$
  and charged particle multiplicity $N_c$), as well as those of
  identified particles will be shown. Finally, the data will be put in
  context with measurements at other beam energies. These systematics
  will then be discussed in terms of possible signatures of the QCD
  phase transition.
\end{abstract}

\section{Introduction}

In order to describe the evolution of ultrarelativistic heavy ion
collisions theorists mainly employ two types of models. On the one hand,
there are cascade-type calculations such as {\sc RQMD}\cite{rqmd},
{\sc ARC}\cite{arc} or, {\sc ART}\cite{art}. Except for the case when
mean field effects are explicitly included into the calculations, these
codes do not contain any 'collectivity'. On the other hand, there are
hydrodynamical models\cite{scheid,rischkeex,rischkeqgp,puersuen,hung}
that do not yet attempt to describe the hadronization phase of
the collision, but provide an insight into the collective expansion of a
nuclear fluid with a given equation of state. Both types of codes have
been successfully applied to describe selected sets of experimental
data. While it is agreed, that the hadronic freezeout obstructs the
'view' into the early stage of the collision some of the models predict
a distinct minimum in the {\em sideward flow}\cite{rischkeex} as a
function of bombarding energy, if a phase transition to the quark gluon
plasma (QGP) is assumed. Another interesting prediction is that as the
bombarding energy is raised, there will be a point ('softest point') at
which the pressure over energy density will reach a minimum at the phase
transition and the system will develop a maximum in the collective 
{\em radial} flow\cite{hung,rischkesoft}.
 
Here, I would like to review some of the recent experimental advances made  
at the AGS in terms of the observation of collective flow. Measurements
of the E877 collaboration, in particular of the reaction plane, will be
presented in some detail and results from other collaborations (E802,
E866, E891) will be used where appropriate.

\section{Event by Event Reconstruction of the Reaction Plane}

The first evidence for the presence of collective motion at AGS energies
(about $10\cdot$AGeV/c) was found in the systematic study of the
transverse energy carried per charged particle as a function of the mass
of the collision system\cite{flowpppa}. As the system size increases the
transverse energy per charged particle increases drastically. Shortly
after this, evidence for {\em sideward flow} was found through a Fourier
analysis of the azimuthal transverse energy distribution
\cite{flowprl}. It was shown, how the 
{\em sideward flow} reaches a maximum for semicentral events and
vanishes for peripheral and the most central collisions. 
In this analysis the reaction plane was not reconstructed on an event by
event basis. 
Having established the {\em sideward flow} effect, it is now possible to
reconstruct the reaction
plane for each event except for the most central events, in which the
reaction plane is not defined and in peripheral events in which not
enough energy is deposited in the calorimeters used in the analysis.
The calorimeters (TCal,PCal - for a description of the detectors and the
methods outlined below see
\cite{flowprc}) are both highly segmented in polar and azimuthal
angle. Each calorimeter cell $j$ measures a fraction $E_T^j$ of the total 
transverse energy $E_T$ at an azimuthal angle $\phi_j$. The
azimuthal angle $\Psi_n^{(i)}$ of the $n$-th moment of the transverse energy
distribution in a pseudorapidity interval $\eta_i$ is given by 

\beqn
\tan \Psi_n^{(i)} = \frac
{\sum_{j}(\pm)E_T^j \sin n\phi_j}
{\sum_{j}(\pm)E_T^j \cos n\phi_j}
\eeqn

where the sign is taken negative (positive) for $\eta$ backward
(forward) of midrapidity. Assuming that the only correlation between
different rapidity 
intervals is due to the initial direction of the impact parameter vector
$\vec{b}$ the angle $\Psi_1^{(j)}$ reflects the orientation of the
reaction plane $\Psi_R$ with some resolution. This resolution can be
measured by a pairwise correlation of the angles $\Psi_1$ from at least
three independent $\eta$ intervals

\beqn
\langle \cos (\Psi_1^{(i)}-\Psi_1^{(j)}) \rangle = \langle \cos
(\Psi_1^{(i)}-\Psi_R)\rangle \langle \cos (\Psi_1^{(j)}-\Psi_R)\rangle.
\eeqn

The azimuthal distribution of any observable $X$ (eg. $E_T,N_c$) with
respect to the reaction plane can be studied by a Fourier decomposition
of the form 

\beqn
v_n'=
\frac
{\langle \sum_{k}X^k  \cos n(\phi_k-\Psi_1^{(i)}) \rangle}
{\langle \sum_{k}X^k  \rangle}.
\eeqn

This will be the subject of the following two sections.
In order to get to the resolution corrected values $v_n$, the measured
Fourier coefficients $v_n'$ need to be unfolded 

\beqn 
v_n = \frac{v_n'}{|\langle \cos n(\Psi_1^{(i)}-\Psi_R) \rangle|}.
\eeqn

The correction factors for the first moment $v_1$ associated with the {\em
sideward flow} and the second moment $v_2$ associated with the {\em
elliptic flow} are shown in Fig.~1.
\begin{center}
  \begin{minipage}[t]{13cm}
    \vskip75mm
    \includegraphics{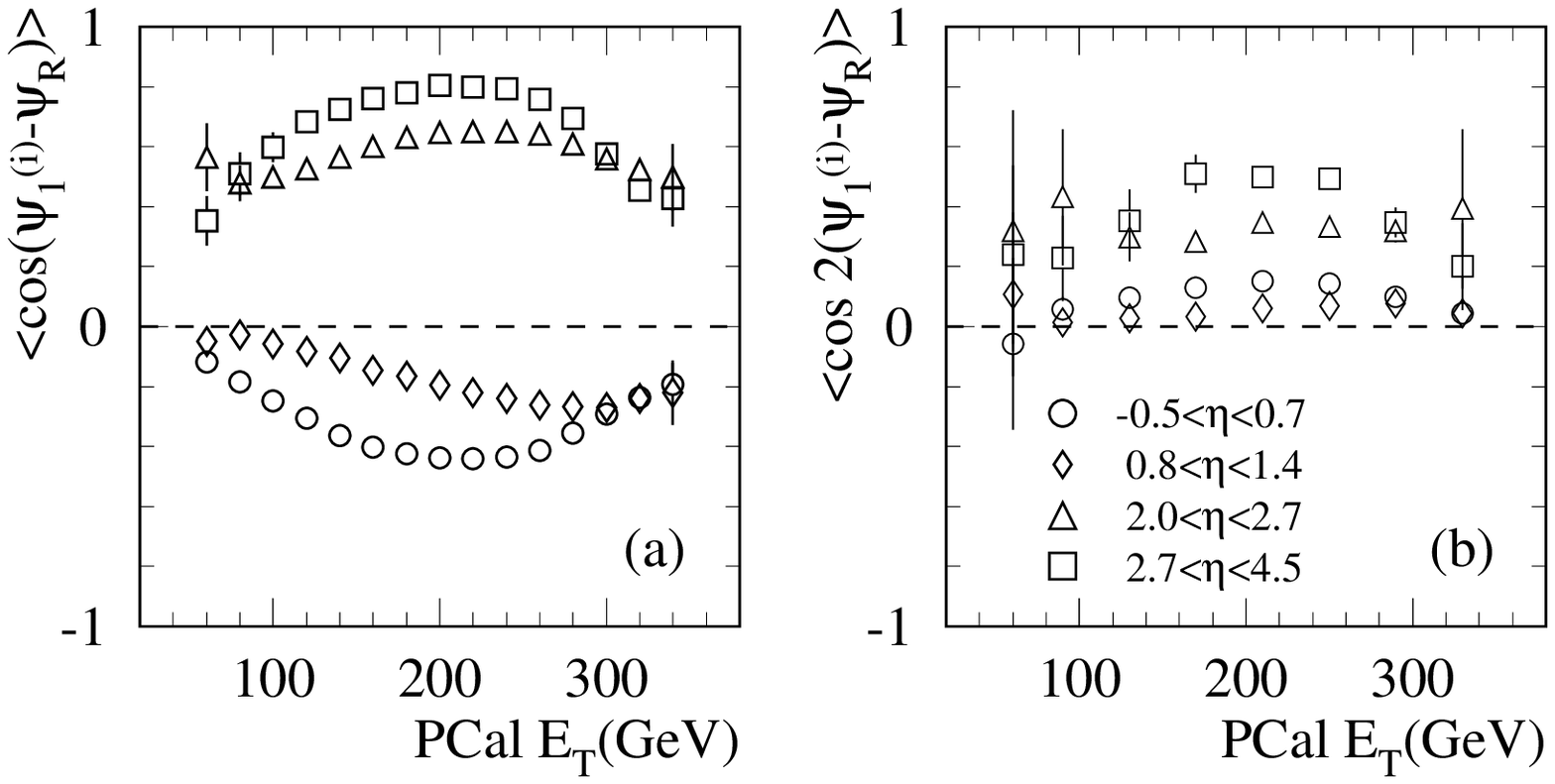}
    \baselineskip=12pt
    {\begin{small}
        Fig.~1. The inverse correction factor for the first moment $v_1$
        (a) and the second moment $v_2$ (b) due to the finite resolution
        in the reaction plane angle $\Psi_R$ for four different bins in
        pseudorapidity as a function of centrality (from E877\cite{flowprc}).
      \end{small}}
  \end{minipage}
\end{center}

\section{Correlation of Global Observables with the Reaction Plane}
Right downstream of the target E877 has a silicon pad multiplicity
detector\cite{flowpppa}. This device consists of two discs each segmented
in polar and azimuthal angle into 512 pads. For the data shown in
Fig.~2 the target calorimeter TCal ($-0.5 < \eta< 0.8$) has been used to
determine the reaction plane angle $\Psi_1$. For some
intermediate centrality bin, where the {\em sideward flow} is maximal 
\cite{flowprl} the charged particle multiplicity is shown as measured with
respect to the reaction plane. In panel (b) three slices through the
double differential distributions plotted in (a) are shown. 
The distributions are peaked in the forward (backward) rapidity slice
$\eta=2.4 (1.1)$ back to back with respect to each other. 
Note, that the reaction plane was determined in an independent

\begin{center}
  \begin{minipage}[t]{13cm}
    \vskip14cm
    \includegraphics{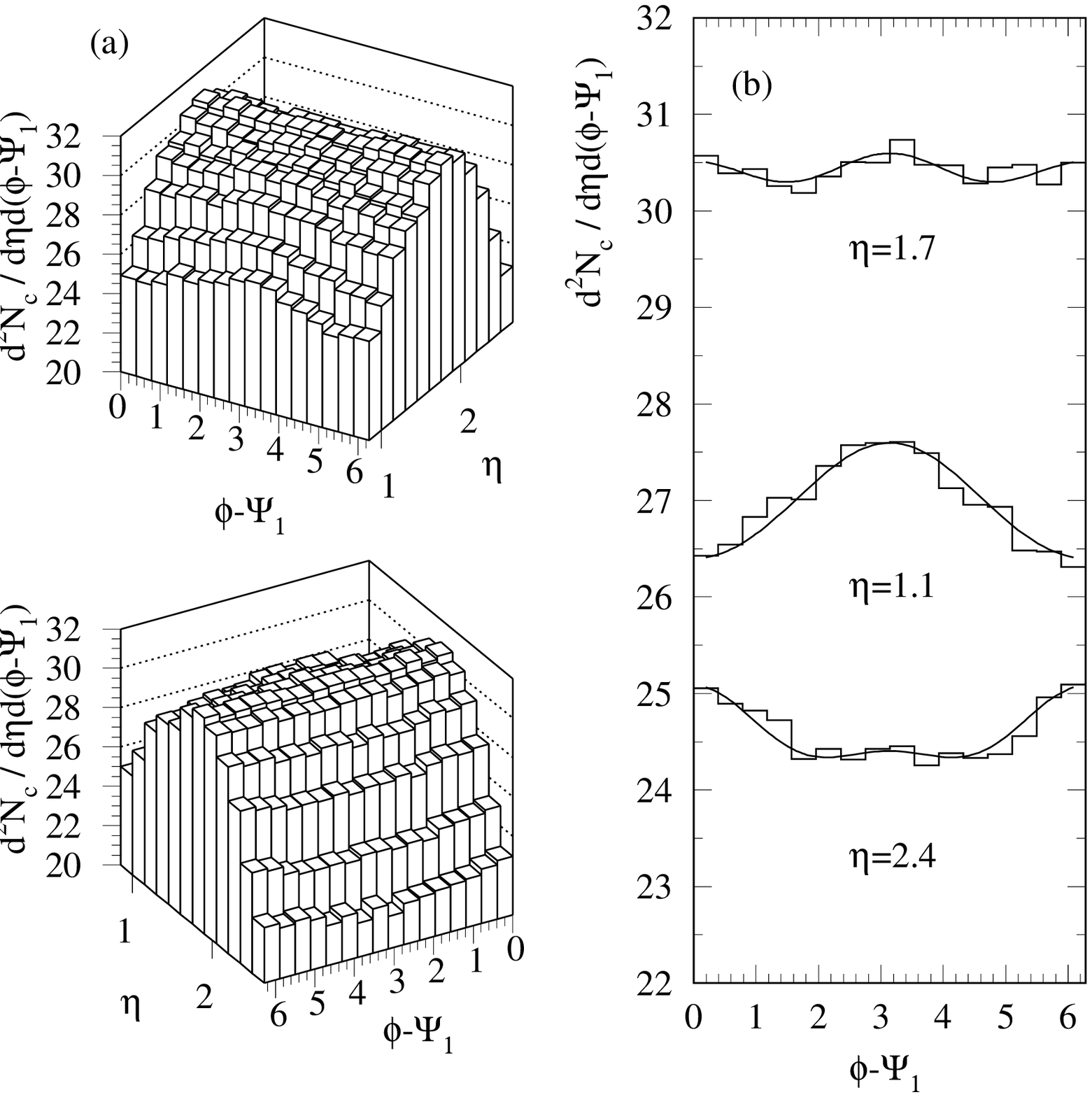}
    \baselineskip=12pt
    {\begin{small}
        Fig.~2. (a) Double-differential charged particle distribution
        for an intermediate centrality bin. Three pseudorapidity slices
        through the distribution are shown in (b) along with fits of
        a Fourier series up to second order (from E877\cite{flowprc}).  
      \end{small}}
  \end{minipage}
\end{center}

way in a different pseudorapidity window. 
In the bin at central rapidity $\eta=1.7$ and to some extent
also in the forward window one can clearly see a nonzero quadrupole
component in the multiplicity distribution. This corresponds to an
elliptic shape of the azimuthal distribution with the major axis in the
reaction plane.

The azimuthal distribution of transverse energy with respect to the
reaction plane has been studied by Fourier decomposition as outlined in
the previous section. Due to the near $4\pi$ calorimetric coverage of
the experiment the full $\eta$ range could be analyzed. The reaction plane
was always determined in a different $\eta$ region than the one analyzed
to avoid autocorrelations. The distributions are shown in
Fig.~3 in windows of increasing centrality ($E_T$) from top left to
bottom right. The solid symbols are the values obtained for the dipole
component $v_1$. One can clearly see how it increases from peripheral to

\begin{center}
  \begin{minipage}[t]{13cm}
    \vskip10cm
    \includegraphics{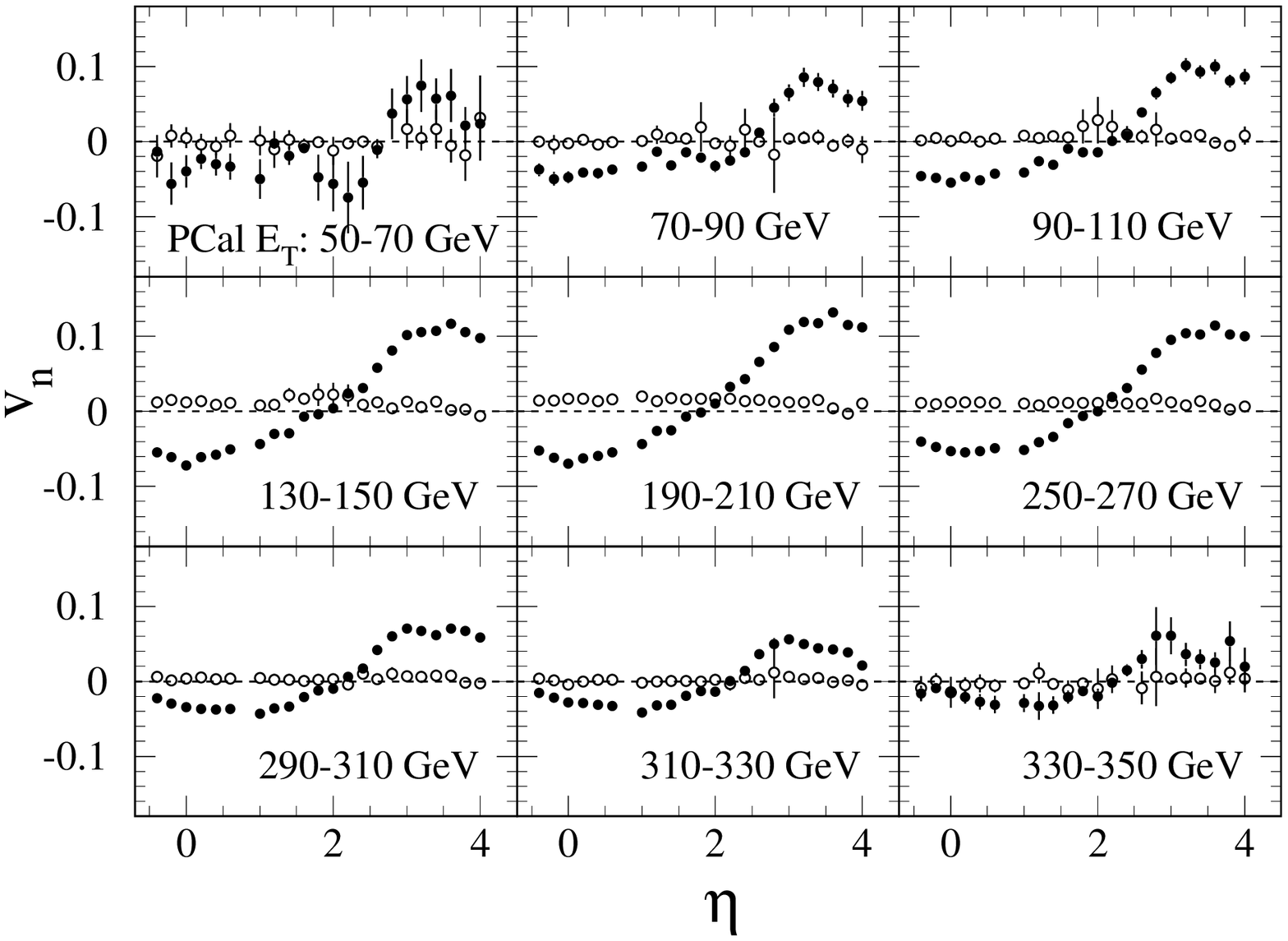}
    \baselineskip=12pt
    {\begin{small}
        Fig.~3. First Fourier coefficients $v_1$ (full symbols) and
        second Fourier coefficients $v_2$ (open symbols) of the azimuthal
        transverse energy distributions versus
        pseudorapidity $\eta$ as a function of centrality (peripheral:
        top left; central: bottom right) (from E877\cite{flowprc}).
      \end{small}}
  \end{minipage}
\end{center}

semicentral collisions and then decreases again for the most central
collisions. The curves are not symmetric about midrapidity since protons
and pions contribute differently for different pseudorapidities. In the
following section this and the fact that the response to the charged
particle multiplicity and transverse energy measurement will be
exploited to disentangle the contributions of nucleons and pions to the
measured {\em sideward flow}.

As was already noted in the distribution of charged particles the
quadrupole component $v_2$ is nonzero. For intermediate centralities
this is the case at all pseudorapidities.
Recently, it has been proposed that both, magnitude and
orientation, of this moment may be sensitive to the pressure developed
early in the collision\cite{sorgep}. There, the magnitude of
the observed {\em elliptic flow} can only be reproduced if mean fields are
included into the calculation (cf. Fig.~6).

\section{Sideward Flow of Identified Particles}

The observed anisotropies in the transverse energy $v_1^{(E_T)}$ and
charged particle $v_1^{(N_c)}$
distributions are a superposition of the contributions due to nucleons
$v_1^n$

\begin{center}
  \begin{minipage}[t]{13cm}
    \vskip10cm
    \includegraphics{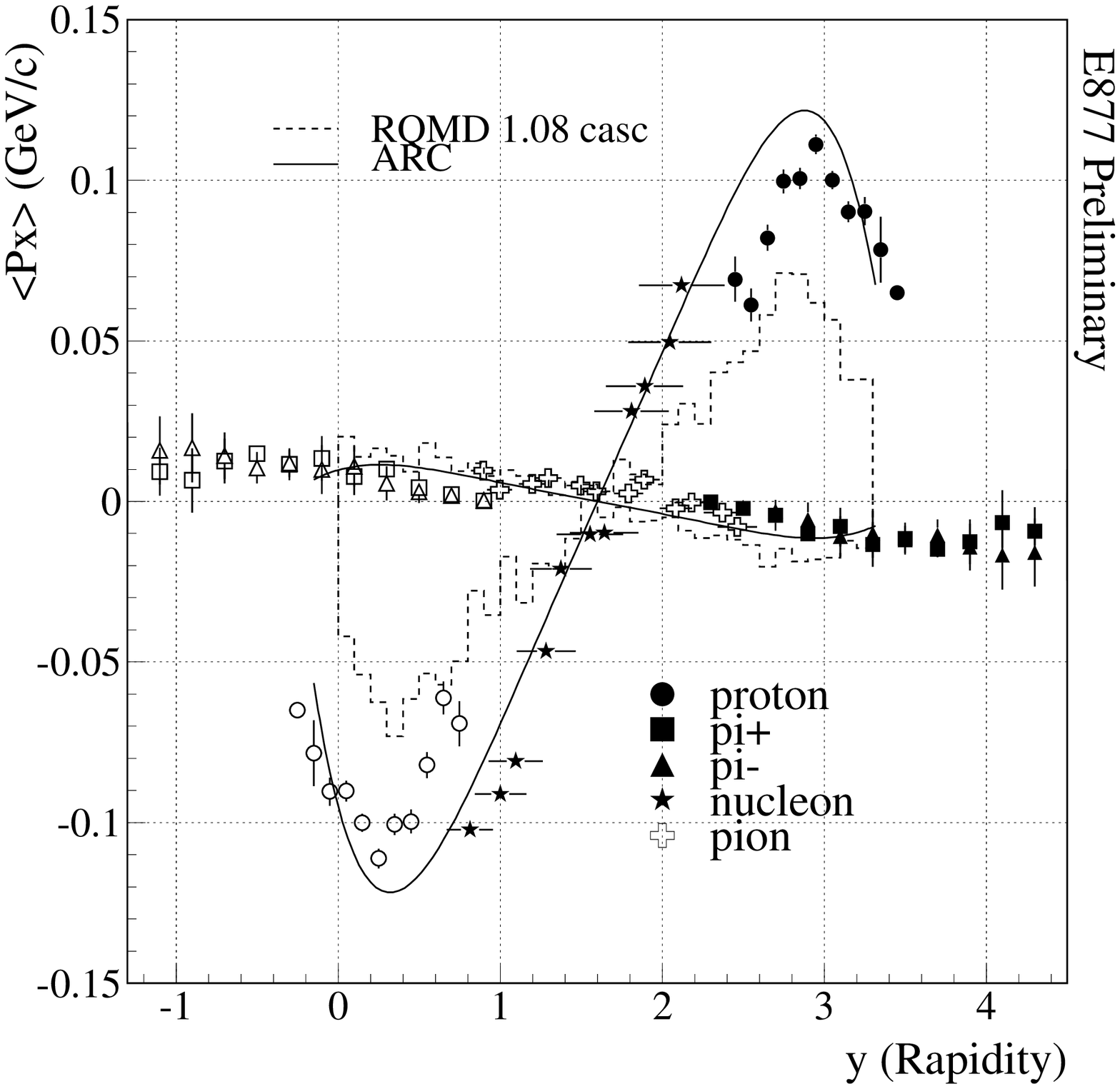}
    \baselineskip=12pt
    {\begin{small}
        Fig.~4. Mean transverse momentum $\langle p_x \rangle$ of
        protons and pions in the reaction plane versus $y$
        (from E877\cite{tkhqm96}) compared to 
        two different model calculations.
      \end{small}}
  \end{minipage}
\end{center}

and pions $v_1^{\pi}$. They can be obtained by solving 

\beqn
v_1^{(N_c)} =  \frac{dN_c^{\pi}/d\eta \; \cdot v_1^{(N_c,\pi)} 
+ dN_c^{n}/d\eta \; \cdot v_1^{(N_c,n)}} 
          {dN_c^{\pi}/d\eta + dN_c^{n}/d\eta },
\eeqn
\beqn
v_1^{(E_T)} = \frac{dE_T^{\pi}/d\eta \; \cdot v_1^{(E_T,\pi)} 
+ dE_T^{n}/d\eta \; \cdot v_1^{(E_T,n)}} 
           {dE_T^{\pi}/d\eta + dE_T^{n}/d\eta }
\eeqn
for $v_1^n$ and $v_1^{\pi}$ with some minimal assumptions (for details
see \cite{flowprc}).  
Multiplying theses coefficients with the measured 
$\langle p_t \rangle $ of protons and pions\cite{akibaqm96,rogerqm96}
one obtains the
average transverse momentum $\langle p_x \rangle$ in the reaction plane
versus rapidity.
The solid stars and open crosses depicted in Fig.~4 are the results of
this deconvolution. 
They are shown along with the analysis of the now
available\cite{flowpid} triple differential cross sections of identified
protons, $\pi^+$, and $\pi^-$ (circles, squares, and triangles). The
histograms show the results of two cascade calculations. {\sc ARC}
with an energy dependent mixture of repulsive and attractive scattering
for the individual baryon collisions (cf.\cite{kahana})
reproduces the data rather well. {\sc RQMD} in its cascade mode
underpredicts the observed flow of protons. With the inclusion of a
repulsive mean field {\sc RQMD} also gives reasonable agreement with the
data.

Experimentally, it is interesting to note that the two pion species show
almost exactly the same magnitude of $\langle p_x \rangle$. If there
were a sizable residual Coulomb interaction with the strongly flowing
protons this would not be expected.

\section{Transverse Radial and Longitudinal Flow of Identified
  Particles}

In an analysis of measured particle ratios, both at the AGS and the SPS
from Si- and S-induced reactions respectively, it was found that the
hadrochemical composition of the final state can be described with the
emission from an equilibrated system\cite{plbs}. The temperatures deduced
were T=120-140\,MeV and 160-170\,MeV respectively
(cf. Fig.~6). Furthermore, it 
was noted that the transverse mass spectra at midrapidity of different
mass particles could only be described, if on top of the thermal energy
the particles were subject to collective transverse {\em radial
  flow}. For the quoted temperatures and a flow profile linear in
radius, the 
average radial expansion velocities $\langle \beta_t \rangle$ were
0.36$\pm$0.03 and 0.27$\pm$0.03 respectively at AGS and SPS. 
For the heaviest collision system \auau{} at the AGS an analysis of data
presented at Quark Matter '96 show that there 
$\langle \beta_t \rangle = 0.45\pm0.03$. 

Similarly, the shape of the rapidity distribution of identified
particles, as shown in Fig.~5, can only be reconciled with emission from a
thermal source, if that source is {\em longitudinally} expanding. The dashed
lines represent emission from a stationary thermal source, while the
solid lines correspond to emission

\begin{center}
  \begin{minipage}[t]{13cm}
    \vskip12cm
    \includegraphics{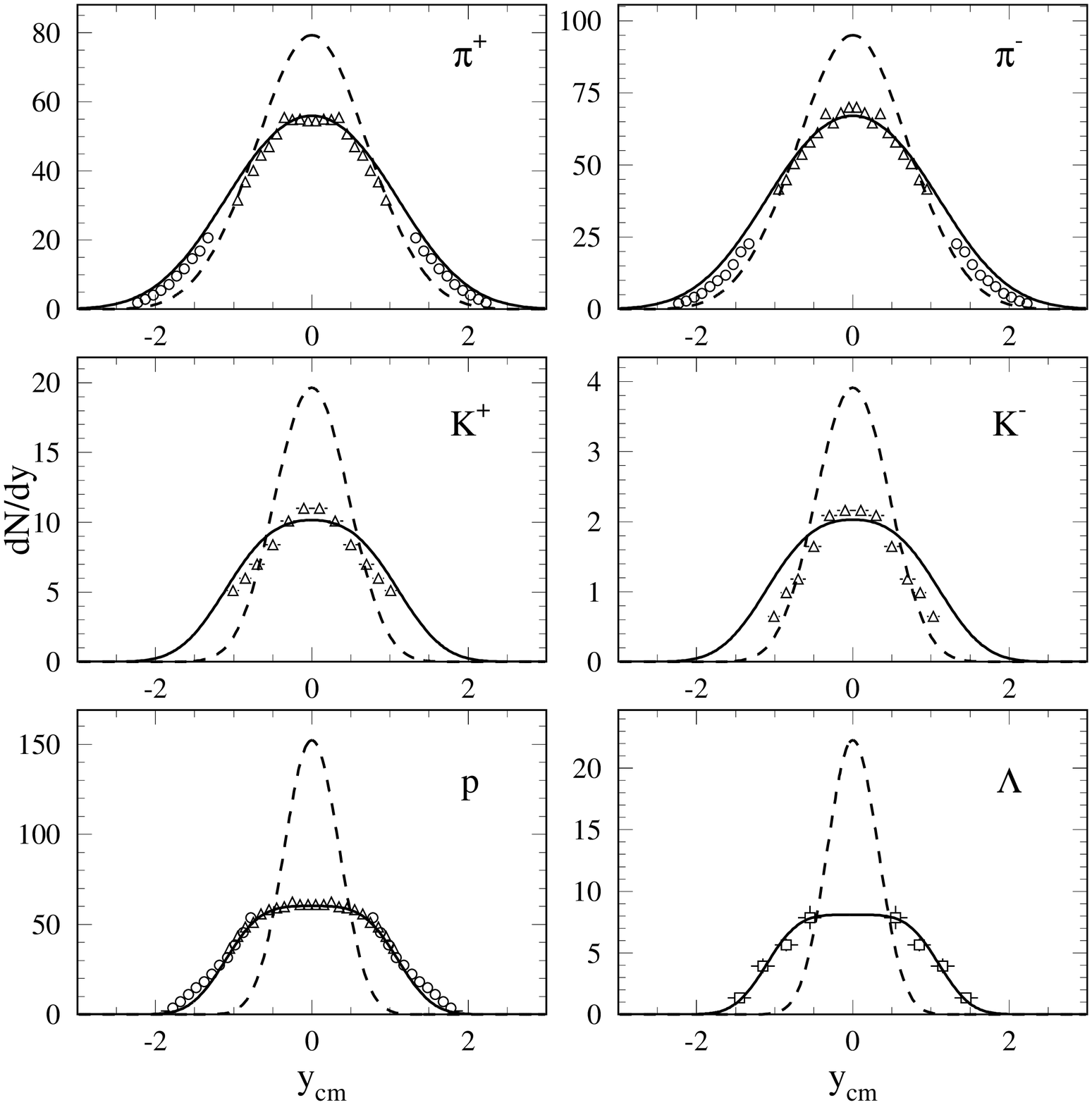}
    \baselineskip=12pt
    {\begin{small}
        Fig.~5. Rapidity distributions of identified particles from
    \auau{} collisions at 11$\cdot$AGeV/c. Data are from E866
    (triangles\cite{akibaqm96}), E877 (circles \cite{rogerqm96}), 
    and E891 
    (squares\cite{lambdas}). See text for a description of the curves.
      \end{small}}
  \end{minipage}
\end{center}

from a thermal source expanding with an average longitudinal velocity
$\langle \beta_l \rangle = 0.5$. This is the 
same average velocity as the one extracted from the relatively small
collision system Si$+$Al at the AGS. All but the $K^-$ spectra are
well described with such an assumption.

\section{Systematics of Flow}

A few interesting things can be learned from a systematic study of the
various types of flow for different collision systems and bombarding
energies. The freezeout temperature rises continuously as the
bombarding energy is increased. Both, at AGS and SPS energies they are
in accordance with the temperatures expected at 
the phase transition between a hot hadron gas and the QGP
\cite{plbs}. This would provide circumstantial evidence for the fact,
that the system may have made the transition to the QGP before
freezeout.

 \begin{center}
  \begin{minipage}[t]{13cm}
    \vskip13cm
    \includegraphics{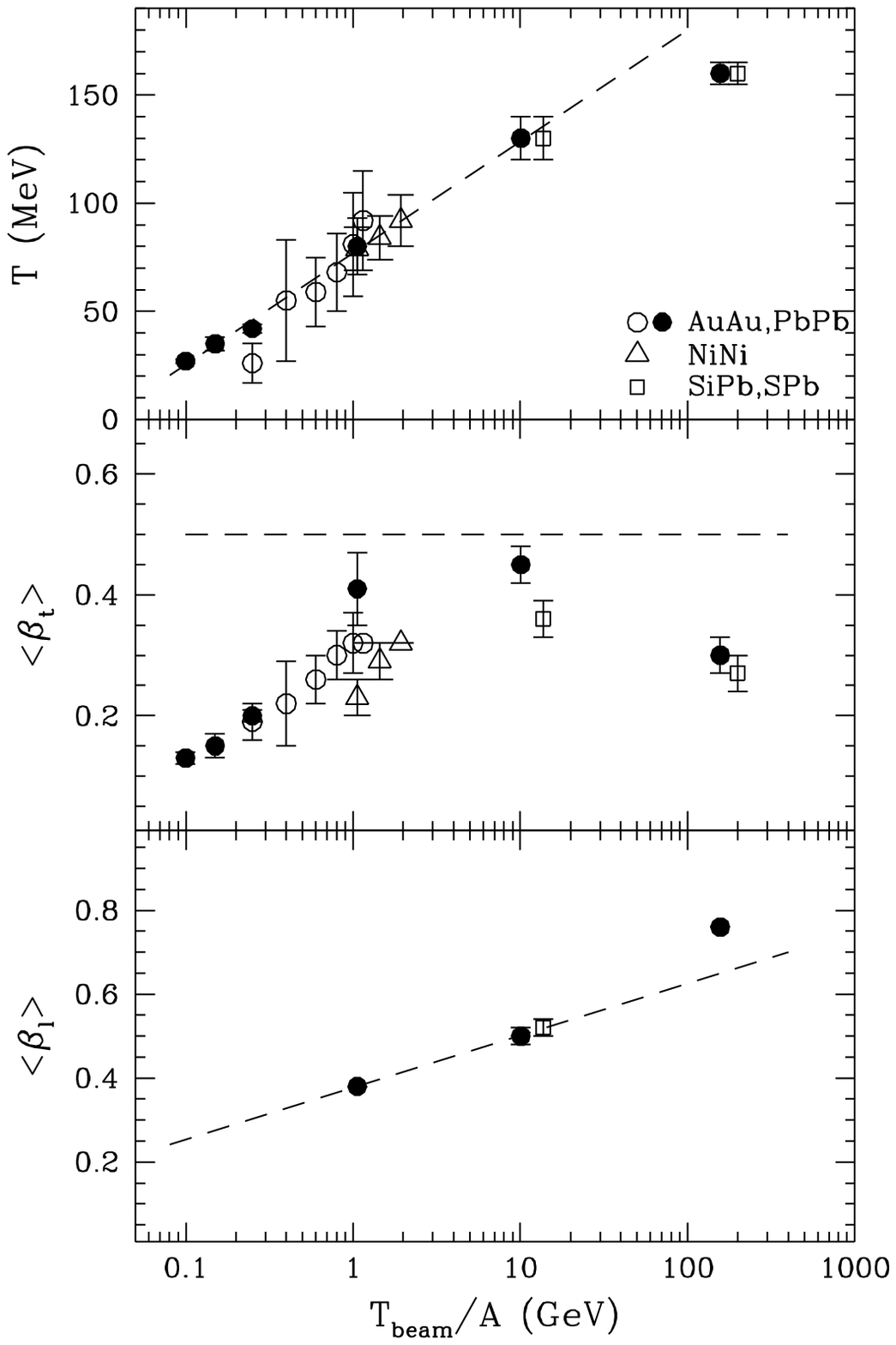}
    \baselineskip=12pt

    \includegraphics{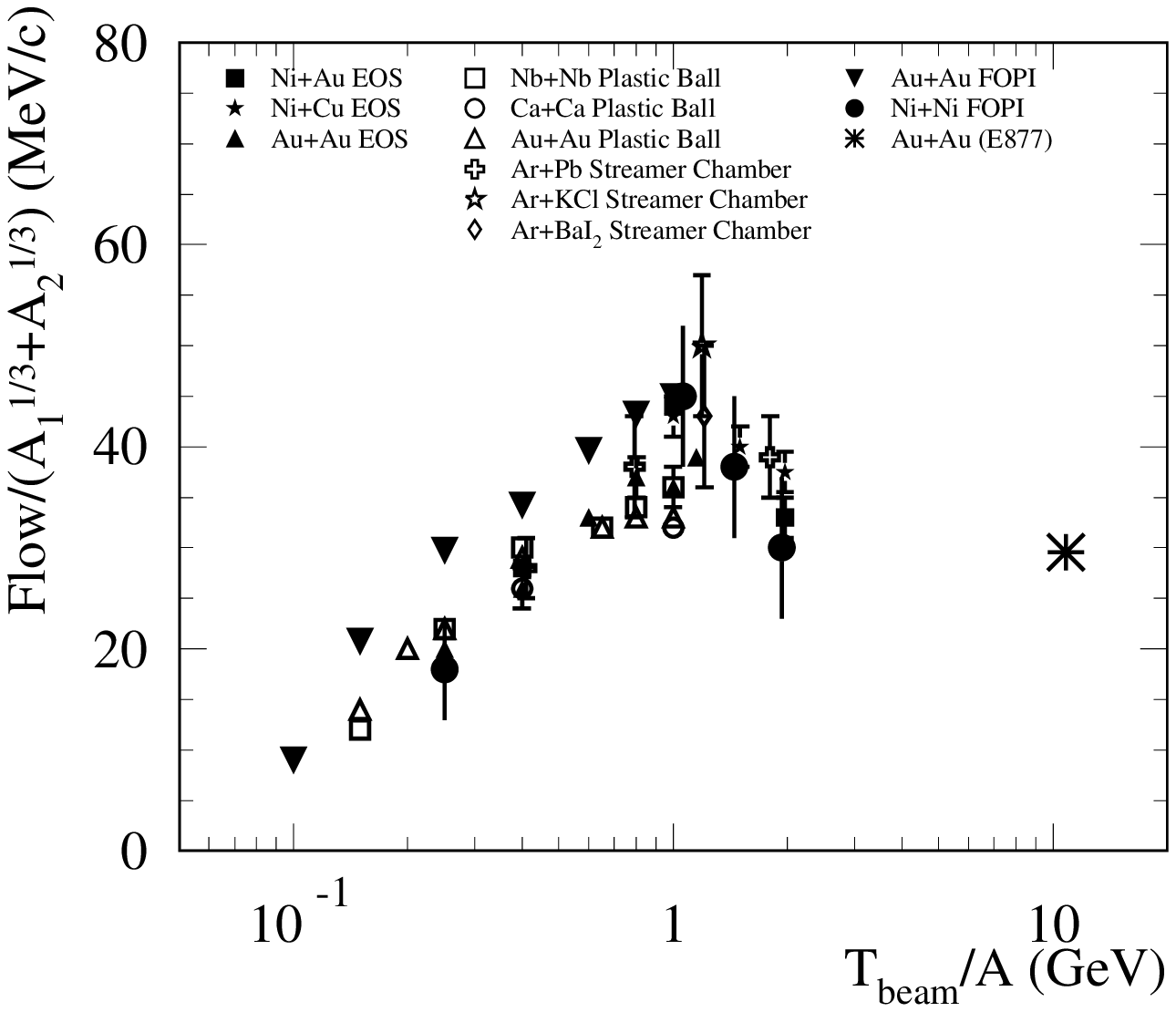}
    \baselineskip=12pt

    \includegraphics{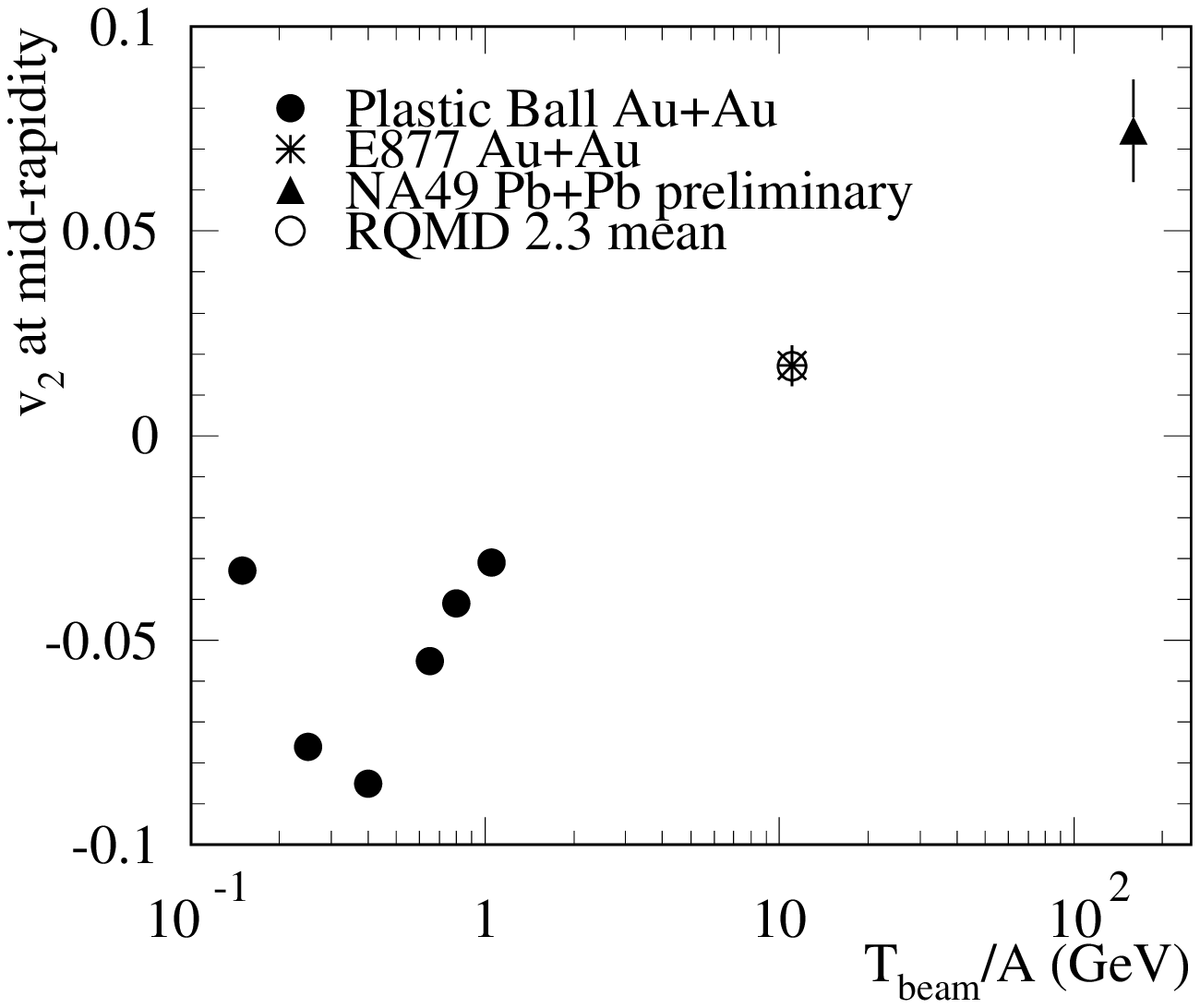}
    \baselineskip=12pt
    {\begin{small}
        Fig.~6. Shown as a function of the beam kinetic energy are  
        (left) temperatures, average transverse
        {\em radial}, and {\em longitudinal} expansion velocities 
        (from\cite{jsqm96}), (right top) {\em sideward flow} 
        (from\cite{nhqm96}), and (right bottom) {\em elliptic flow} 
        (data from\cite{flowpid,plaball,wienoldqm96}). 
        See text for details.
      \end{small}}
  \end{minipage}
\end{center}

At energies between SIS and SPS a maximum in the transverse {\em radial
  flow} is reached. A maximum is expected, where the lifetime of the
system is largest such that a high degree of collectivity in the
subsequent expansion can be achieved - one of the predictions of
hydrodynamics with a phase transition\cite{hung}.

The {\em longitudinal flow} velocities increase faster than
logarithmically between AGS and SPS energies. This is expected, if
stopping seizes to be complete. 

The {\em sideward flow} $F$ (top right) in this representation is
defined as $F=(d\langle p_x/A \rangle /dy')/(A_1^{1/3}+A_2^{1/3})$
where $y'=y/y_{beam}$ and $A_{1,2}$ are the masses of projectile and
target in order to be able to compare to asymmetric systems (see
\cite{nhqm96}. The E877 data was added to the systematics). The data
seem to level off between SIS and AGS energies.
One-fluid hydrodynamic calculations\cite{rischkeex} predict a vanishing
of this quantity between the two energies. Data are not yet conclusive
here.

{\em Elliptic flow} has first been discovered by the Plastic Ball at the
Bevalac\cite{plaball} and was coined 'squeezeout' because there the
major axis was perpendicular to
the reaction plane, reminiscent of the shadowing due to the projectile
and target nucleons. In raising the beam energy the sign of $v_2$
changes and is now more sensitive to the high initial pressure in the
collision\cite{sorgep}. Whether this trend continues at the SPS as shown
in the figure is not completely clear yet, since
the data from NA49\cite{wienoldqm96} does not allow for the
reconstruction of the reaction plane and therefore the sign of $v_2$ has
not been verified. Also, the magnitude depends on the assumption that
pions contribute mostly to the measured signal and that for them
transverse momentum and transverse energy are similar.

\section{Outlook}

With the advent of very detailed datasets, it should finally be possible
to determine some of the fundamental thermal and hydrodynamical
properties of highly excited nuclear matter and to discern some of the
models that try to describe it. At SPS energies first
direct hints of the elusive QGP have been reported through the
measurement of dielectrons\cite{ad} and dimuons\cite{mg}. On the
hadronic side this is corroborated by the findings of the freezeout
temperature at the phase transition line. These signals should be
largest, when the largest system with highest energy and baryon density
is formed. In order to determine this point, the EOS collaboration is now
exploring the bombarding energy region between SIS and AGS energies and
the CERN experiments will prepare for a run at the lowest attainable
energy at the SPS to get data in the region above AGS energies. 

\section{Acknowledgements}

In the presentation of the results on {\em sideward flow}, I profited
especially from contributions of S.~Voloshin and W.C.~Chang. 
In the discussion of the systematics of the various types of flow, I
enjoyed discussions with N.~Herrmann and J.~Stachel.

\end{document}